\documentstyle[epsf,longtable]{aipproc}

\def\ra{\rightarrow}

\catcode`@=11
\long\def\@makefntext#1{\parindent 0pt\hsize\columnwidth\parskip0pt\relax
\footnotesize\baselineskip12pt\def\strut{\vrule width0pt height0pt depth1.75pt\relax}%
\mbox{$\m@th^{\@thefnmark}$\hspace*{3pt}}#1}
\catcode`@=12

\begin{document}
\pagestyle{plain}

\font\fortssbx=cmssbx10 scaled \magstep2
\hbox to \hsize{{  }  
\hfill
\vtop{\hbox{\bf UH-511-960-00}
      \hbox{\bf March 2000}}}

\title{Do Neutrinos Decay?\footnote{Presented at the {\it 5th
International Conference on Physics
Potential \& Development of $\mu^+\mu^-$ Colliders}, San Francisco, December 1999.}}

\author{S. Pakvasa}

\address{\vglue-3ex
Department of Physics \& Astronomy  \\
University of Hawaii  \\ 
Honolulu, HI 96822\vglue-3ex}

\maketitle

\begin{abstract}
If neutrinos have masses and mixings, in general some will decay.  In
models  with new particles and new couplings, some decay modes can be
fast enough to be of phenomenological interest.
\end{abstract}
\thispagestyle{empty}

\section{Introduction}

It is generally agreed that most probably neutrinos have non-zero masses
and non-trivial mixings.  This belief is based primarily on the evidence
for neutrino mixings and oscillations from the data on atmospheric
neutrinos and on solar neutrinos\cite{barger}.

If this is true, then in general the heavier neutrinos are expected to
decay into the lighter ones via flavor changing processes.  The only
questions are (a) whether the lifetimes are short enough to be
phenomenologically interesting (or are they too long?) and (b) what are
the dominant decay modes.

Throughout the following discussion, to be specific, I will assume that
the neutrino masses are at most of order of a few eV.  There are
interesting things to be done with heavier decaying neutrinos 
which we will not discuss here today\cite{kawasaki}.

\section{Radiative Decays}

For eV neutrinos, the only radiative decay modes possible are $\nu_i \ra
\nu_j + \gamma$.  They can occur at one loop level in SM (Standard
Model).  The decay rate is given by\cite{petcov}:
\begin{equation}
\Gamma =
\frac{9}{16} \
\frac{\alpha}{\pi} \
\frac{G^2_F}{128 \pi^3} \
\frac{\left ( \delta m_{ij}^2 \right )^3}{m_i}  
\left | \sum_{\alpha} \ U_{i \alpha}^* \
U_{\alpha j} \
 \left (
\frac{m_\alpha ^2}{m^2_W} \right ) \right |^2
\end{equation}
where $\delta m^2_{ij} = m^2_i - m^2_j$ and $\alpha$ runs over $e, \mu$ and
$\tau$.  When $m_i \gg m_j, m_i \sim O (eV)$ and for maximal mixing
$(4U^{*2}_{i \alpha} U^2_{\alpha j}) \sim O(1)$ and $(\alpha \cong \tau)$
one obtains for $\Gamma$
\begin{equation}
\Gamma_{SM} \sim 10^{-45} \ sec^{-1}
\end{equation}
which is far too small to be interesting.  If the dominant contribution
is from electron (rather than $\tau$) then there is enhancement in
matter (due to the electrons present) by a factor\cite{nieves}
\begin{equation}
10^{24} \left (
\frac{\rho_e}{10^{24} (cc)^{-1}} \right )
\left (
\frac{1eV}{m_i} \right )^4  \ F
\end{equation}
where the function $F$ tends to $4m_i/E_i$ for relativistic $\nu'_is_i$
and this enhancement can be as large as $10^{16}$.

The decay mode $\nu_i \ra \nu_j + \gamma$ comes from an effective
coupling which can be written as:
\begin{equation}
\left (
\frac{e}{m_i + m_j}
\right )
\bar{\psi}_j \sigma_{\mu \nu} (C + D \gamma_5) \psi_i \ F{_\mu \nu} 
\end{equation}
Let us define $k_{ij}$ as
\begin{eqnarray}
k_{ij} & = &
\left (
\frac{e}{m_i + m_j}\right ) \sqrt{\mid C \mid ^2  + \mid D \mid^2} \\ \nonumber
& \equiv &  k_0 \mu_B  \\ \nonumber
\end{eqnarray}
where $\mu_B = e/2m_e$.
Since the experimental bounds on $\mu_{\nu i}$, the magnetic moments of
neutrinos, come from reactions such as $\nu_e e \ra e ``\nu''$ which are
not sensitive to the final state neutrinos; the bounds apply to both
diagonal as well as transition magnetic moments and so can be used to limit
$k^i_0$ and the corresponding lifetimes.  The current bounds are\cite{particle}:
\begin{eqnarray}
k^e_0 \ < \ 10^{-10} \\ \nonumber
k^\mu_0 \ < \ 7.4.10^{-10} \\ \nonumber
k^\tau_0 \ < \ 5.4.10^{-7}
\end{eqnarray}
For $m_i \gg m_j$, the decay rate for $\nu_i \ra \nu_j + \gamma$ is
given by
\begin{equation}
\Gamma = \frac{\alpha}{2 m_e^2} \ m_i^3 \ k_0^2
\end{equation}
This, in turn, gives indirect bounds on radiative decay lifetimes for
$O(eV)$ neutrinos of:
\begin{eqnarray}
\tau_{\nu_{e}} \ > \ 5.10^{18} \ \mbox{sec}  \\ \nonumber
\tau_{\nu_{\mu}} \ > \ 5.10^{16} \ \mbox{sec}  \\ \nonumber
\tau_{\nu_{\tau}} \ > \ 2.10^{11} \ \mbox{sec} 
\end{eqnarray}

There is one caveat in deducing these bounds.  Namely, the form factors
C and D are evaluated at $q^2 \sim O (eV^2)$ in the decay matrix
elements whereas in the scattering from which the bounds are derived,
they are evaluated at $q^2 \sim O (MeV^2)$.  Thus, some extrapolation is
necessary.  It can be argued that, barring some bizarre behaviour, this
is justified\cite{frere}.

There are other bounds on radiative lifetimes of neutrinos which are all
based on non-observation of the final state $\gamma$-ray.  The most
direct observational bounds from reactor and accelerators are\cite{particle}:
\begin{eqnarray}
\tau (\nu_e)  \ > 300 \ s \\ \nonumber
\tau (\nu_\mu) \ > 15.4 \ s 
\end{eqnarray}
There are other more indirect bounds\cite{particle}. 
From cosmology:
\begin{equation}
\tau > 2.10^{21}  \ s.
\end{equation}
From x-ray and $\gamma$-ray fluxes:
\begin{equation}
\tau > 7.10^{9}  \ s.
\end{equation}
Here we have set $m_i \sim O (eV)$.
All these bounds depend on assuming that $m_i \gg m_j$ in the mode
$\nu_i \ra \nu_j + \gamma$.  The bounds are not valid if
there is near degeneracy; since in this case the $\gamma$-ray can be soft.
\section{The Sciama Model}

Dennis Sciama proposed an intriguing and imaginative scenario for
radiative decay of a neutrino about 10 years ago\cite{sciama}.  It went thru several
metamorphoses; starting out as a decay mode $\nu_\tau \ra \nu_\mu +
\gamma$ and ending up most recently as $\nu_{st} \ra \nu'_{st} +
\gamma$\cite{mohapatra}.

The proposal was this:  (a) suppose the decay $\nu_\alpha \ra \nu_\beta
+ \gamma$ takes place, (b) the mass of $\nu_\alpha (m_{\nu \alpha})$ is
$\sim 27.4$ eV, and $m_{\nu_{\alpha}} \gg  m_{\nu_{\beta}},$ (c) thus
$E_\gamma \approx \ 13.7$ eV for a slow moving $\nu_\alpha$; (d) the lifetime by
this mode for $\nu_\alpha$ is  about $2.10^{23}$ sec.  These properties
are consistent  with all the bounds.  The purpose was to account
for the anomalous amount of ionised $H$ in interstellar space.  The
energy of the decay photon is  just right to ionise hyrogen; and the
lifetime is chosen so as to provide the required amount of ionization.

After a lengthy innings (of 10 years), it appears that the Sciama model
is finally ruled out.  The most recent evidence against it is 
the non-observation of the line at 911 \AA (corresponding to 13.7 eV) with the predicted
intensity\cite{bowyer}. 

\section{Invisible Decays}

A decay mode with essentially invisible final states which does not
involve any new particles is the three body neutrino decay mode.  The
decay is $\nu_i \ra \nu_j \nu_j \bar{\nu}_j$.  This decay, like the
radiative mode, can occur at one loop level in SM.  With a mass pattern
$m_i \gg m_j$ the decay rate can be written as
\begin{equation}
\Gamma = \frac{\epsilon^2 \ G^2_F \ m_j^5}{192 \pi^3}
\end{equation}
In the SM at one loop level, with the internal $\tau$ dominating, the
value of $\epsilon^2$ is given by\cite{lee}
\begin{equation}
\epsilon_{SM}^2 = \frac{3}{16} \left (
\frac{\alpha}{\pi} \right )^2
\left ( \frac{m_\tau}{m_W} \right )^4
\left\{
\ell n
\left (
\frac{m_\tau^2}{m_w^2} \right ) \right \}^2
\left ( U_{\tau j} \ U^*_{\tau i} \right )^2
\end{equation}

With maximal mixing $\epsilon^2_{SM} \approx 3.10^{-12}$.  Even if
$\epsilon$ were as large as 1 with new physics contributions; it only gives
a value for $\Gamma$ of $5.10^{-35} \ sec^{-1}$.  Hence, this decay mode
will not yield decay rates large enough to be of interest.  It is worth
noting that the current experimental bound on $\epsilon$ is quite poor:  
$\epsilon  < O(100)$\cite{bilenky}.
\section{$\nu_{\alpha_{L}} \ra \nu_{\beta_{L}} + \chi$}

If the only new particle introduced is an I=0, L=0, J=0, massless
particle such as a Goldstone boson (familon), then a new decay mode
is possible:
\begin{equation}
\nu_{\alpha_{L}} \ra  \nu_{\beta_{L}} + \chi
\end{equation}
which arises  from a coupling of the form
\begin{equation}
g_p \bar{\psi}_{\beta_{L}}  \gamma_\mu \psi_{\alpha_{L}} \partial_\mu \chi
\end{equation}
with a decay rate
\begin{equation}
\Gamma =
\frac{g^2_p m^3_\alpha}{16  \pi}
\end{equation}
$SU(2)_L$ symmetry predicts a similar coupling for the charged leptons
with decay mode $\ell_\alpha \ra \ell_\beta + \chi$.  Thus the
$\nu_\alpha$ lifetime is related to the B.R. $(\ell_\alpha \ra \ell_\beta
+ \chi)$ by
\begin{equation}
\tau_{\nu_{\alpha}} =
\frac{\tau_{\ell \alpha}}{B.R. (\ell_\alpha \ra \ell_\beta + \chi)}
\left (
\frac{m _{\nu_{\alpha}}}{m_{\ell_{\alpha}}} \right )^{-3}
\end{equation}
The current bounds on $\mu$ and $\tau$ branching ratios are\cite{particle,jodidio}
\begin{eqnarray}
B.R. (\mu \ra e \chi) \ < 2.10^{-6} \\ \nonumber
B.R. (\tau \ra \mu \chi) \ < 7.10^{-6}
\end{eqnarray}
and lead to
\begin{eqnarray}
\tau_{\nu_{\mu}} > 10^{24} \ s \\ \nonumber
\tau_{\nu_{\tau}} > 10^{20} \ s.
\end{eqnarray}
Coupling to a scalar field with $I=1/2$, $L=0$ would have the same constraint as 
above; and in addition would require fine tuning to
avoid mixing with the SM Higgs.

\section{Fast Invisible Decays}

The only possibility for fast invisible decays of neutrinos seems to lie
with Majoron models.  There are two classes of models; the I=1
Gelmini-Roncadelli\cite{gelmini} majoron and the I=0 Chikasige-Mohapatra-Peccei\cite{chikasige}
majoron.  In general, one can choose the majoron to be a mixture of the
two; furthermore the coupling can be to flavor as well as sterile
neutrinos.  The effective interaction is of the form:
\begin{equation}
g_\alpha \bar{\nu}_\beta^c \nu_\alpha \ J
\end{equation}
giving rise to decay:
\begin{equation}
\nu_\alpha \ra \bar{\nu}_\beta \ + J
\end{equation}
where $J$ is a massless $J=0 \ L=2$ particle; $\nu_\alpha$ and $\nu_\beta$
are mass eigenstates which may be mixtures of flavor and sterile
neutrinos.  Models of this kind which can give rise to fast neutrino
decays and satisfy the bounds below have been discussed by Valle,
Joshipura and others\cite{valle}.  These models are unconstrained by $\mu$ and
$\tau$ decays which do not arise due to the $\Delta L=2$ nature of the
coupling.  The I=I coupling is constrained by the bound on the invisible
$Z$ width\cite{kawasaki}; and requires that the Majoron be a mixture of I=1 and I=0.
The couplings of $\nu_\mu$ and $\nu_e \ (g_\mu$ and $g_e)$ are
constrained by the limits on multi-body $\pi, k$ decays ${\pi \ra \mu
\nu \nu  \nu}$ and $K \ra \mu \nu \nu \nu$ and on $\mu-e$ university
violation in $\pi$ and K decays\cite{barger1}.

Granting that models with fast, invisible decays of neutrinos can be
constructed, can such decay modes be responsible for any observed neutrino anomaly?

 We assume a
component of $\nu_\alpha,$ i.e., $\nu_2$, to be the only unstable state,
with a rest-frame lifetime $\tau_0$, and we assume two flavor mixing,
for simplicity:
\begin{equation}
\nu_\alpha = cos \theta \nu_2 \ + sin \theta \nu_1
\end{equation}
with $m_2 > m_1$.  From Eq. (2) with an unstable $\nu_2$, the $\nu_\alpha$
survival probability is
\begin{eqnarray}
P_{\alpha \alpha} &=& sin^4 \theta \ + cos^4 \theta {\rm exp} (-\alpha L/E)
            \\ \nonumber
&+& 2 sin^2 \theta cos^2 \theta {\rm exp} (-\alpha L/2E)
            cos (\delta m^2 L/2E),
\end{eqnarray}
where $\delta m^2 = m^2_2 - m_1^2$ and $\alpha = m_2/ \tau_0$.
Since we are attempting to explain neutrino data without oscillations
there are two appropriate limits of interest.  One is when the $\delta
m^2$ is so large that the cosine term averages to 0.  Then the survival
probability becomes
\begin{equation}
P_{\mu\mu} = sin^4 \theta \ + cos^4 \theta {\rm exp} (-\alpha L/E)
\end{equation}
Let this be called decay scenario A.  The other possibility is when
$\delta m^2$ is so small that the cosine term is 1, leading to a
survival probability of 
\begin{equation}
P_{\mu \mu} = (sin^2 \theta + cos^2 \theta {\rm exp} (-\alpha L/2E))^2
\end{equation}
corresponding to decay scenario B.  Decay models for both kinds of
scenarios can be constructed; although they require fine tuning and are
not particularly elegant.

The possibility of solar neutrinos decaying to explain the discrepancy
is a very old suggestion\cite{pakvasa}. 
The most recent analysis of the current solar neutrino data finds that
no good fit can be found:
$U_{ei} \approx 0.6$ and $\tau_\nu$ (E=10
MeV) $\sim 6$ to 27 sec. come closest\cite{acker}. 
The fits become acceptable only if the suppression
of the solar neutrinos is energy independent as proposed by several
authors\cite{harrison} (which is possible if
the Homestake data are excluded from the fit).  The above conclusions are
valid for both the decay scenarios A as well as B.

For atmospheric neutrinos, assuming neutrino decay scenario A, it was found that it 
is possible to choose $\theta$ and $\alpha$ to provide a
good fit to the Super-Kamiokande L/E distributions of $\nu_\mu$ events and
$\nu_\mu/\nu_e$ event ratio\cite{barger2}.  
The best-fit values of the two parameters are $cos^2 \theta
\sim 0.87$   and $\alpha \sim 1 GeV/D_E,$ where
$D_E=12800$ km is the diameter of the Earth.  
This best-fit $\alpha$ value corresponds
to a rest-frame $\nu_2$ lifetime of
\begin{equation}
\tau_0 = m_2/\alpha \sim 
\frac{m_2}{(1 eV)} \times 10^{-10} s.
\end{equation}
However, it was then shown that the fit to the higher energy events in
Super-K (especially the upcoming muons) is quite poor\cite{lipari}.
The reason that the inclusion of high energy
upcoming muon events makes the fits poorer is very simple. 
In the decay A
scenario, due to time dilation, the decay is suppressed at high energy 
and there is
hardly any depletion of $\nu_\mu's$, contrary to observation.

Turning to decay scenario B, consider the following possibility\cite{barger3}.  The
three states $\nu_\mu, \nu_\tau, \nu_s$ (where $\nu_s$ is a
sterile neutrino) are related to the mass eigenstates $\nu_2 \nu_3
\nu_4$ by the approximate mixing matrix.
\begin{equation}
\left( \begin{array}{c} \nu_\mu\\ \nu_\tau\\ \nu_s \end{array} \right) =
\left( \begin{array}{ccc}  \cos\theta& \sin\theta& 0\\
                          -\sin\theta& \cos\theta& 0\\
                           0& 0& 1
\end{array} \right)
\left( \begin{array}{c} \nu_2\\ \nu_3\\ \nu_4 \end{array} \right)
\label{eq:mixing}
\end{equation}
and the decay is $\nu_2 \to \bar\nu_4 + J$. The electron neutrino,
which we identify with $\nu_1$, cannot mix very much with the other
three because of the more stringent bounds on its couplings\cite{barger1},
and thus our preferred solution for solar neutrinos would be
small angle matter oscillations.

In this case the $\delta m_{23}^2$ in Eq.~(23) is not
related to the $\delta m_{24}^2$ in the decay, and can be very small,
say $ < 10^{-4}
\rm\, eV^2$ (to ensure that oscillations play no role in the atmospheric
neutrinos). Then the oscillating term is 1 and $P(\nu_\mu\to
\nu_\mu)$ is given by Eq. (25).

In order to compare the  predictions of this model with the standard
$\nu_\mu \leftrightarrow \nu_\tau$ oscillation model,
we have  calculated  with Monte Carlo methods   the  event  rates
for contained, semi-contained  and upward-going  (passing and stopping)
muons  in the Super-K  detector,  in the  absence of `new  physics', and
modifying the muon neutrino  flux   according
to the   decay or  oscillation  probabilities   discussed  above.
We have then  compared our  predictions  with the SuperK data
\cite{fukuda},  calculating  a $\chi^2$  to
quantify the agreement (or  disagreement)  between  data and
calculations.
In    performing  our  fit  (see Ref.~\cite{barger3} for details)
we  do not take into account any  systematic  uncertainty, but we allow
the  absolute flux normalization to vary as  a free
parameter  $\beta$.

The decay  model of Equation (24) above gives a very good fit
with a minimum $\chi^2 = 33.7$ (32 d.o.f.)
for the choice  of  parameters
\begin{equation}
\tau_\nu/m_\nu = 63\rm~km/GeV,
\ \cos^2 \theta = 0.30
\end{equation}
and normalization $\beta = 1.17$.

In Fig.~1  we  compare  the
best fits of the two  models  considered
(oscillations  and  decay)  with the
SuperK  data.
In the figure we show
(as  data points  with statistical error bars)
the ratios between the SuperK data and the Monte Carlo  predictions
calculated  in the  absence of oscillations or other
form of `new physics' beyond the standard model.
In the six  panels  we  show   separately  the  data
on $e$-like and  $\mu$-like events in the sub-GeV and multi-GeV
samples,   and  on   stopping and passing
upward-going muon events.
The   solid (dashed) histograms
correspond to  the  best fits for the decay  model
($\nu_\mu \leftrightarrow \nu_\tau$ oscillations).
One  can  see  that the  best fits  of the two  models
are  of comparable  quality.
The reason  for the similarity  of the results  obtained
in  the two  models  can be understood by looking  at
Fig.~2, where  we show
the survival probability $P(\nu_\mu \to \nu_\mu)$
of muon neutrinos   as  a  function
of $L/E_\nu$ for  the  two  models   using the
best  fit  parameters.
In the case  of the neutrino  decay model   (thick  curve)
the probability   $P(\nu_\mu \to \nu_\mu)$
monotonically  decreases   from    unity  to  an  asymptotic  value
$\sin^4 \theta \simeq  0.49$.
In the case of  oscillations the  probability  has  a sinusoidal
behaviour  in $L/E_\nu$.  The  two  functional    forms
seem     very different;  however,  taking  into  account  the
resolution in $L/E_\nu$,  the  two  forms
are  hardly  distinguishable.
In fact, in the    large $L/E_\nu$    region, the oscillations
are  averaged  out  and the survival  probability there
can  be  well  approximated  with 0.5  (for  maximal  mixing).
In  the region  of  small  $L/E_\nu$  both probabilities  approach
unity.
In the region $L/E_\nu$ around  400~km/GeV, where  the  probability for the
neutrino oscillation model  has the first  minimum,
the  two  curves are  most  easily  distinguishable, at least in
principle.

\medskip\noindent{Decay Model}

There are two decay possibilities that can be considered: (a)~$\nu_2$ decays to
$\bar\nu_4$  which is dominantly $\nu_s$ with $\nu_2$ and $\nu_3$ mixtures of
$\nu_\mu$ and $\nu_\tau$, as in Eq.~(\ref{eq:mixing}), and
(b)~$\nu_2$ decays into $\bar\nu_4$ which is dominantly $\bar\nu_\tau$ and
$\nu_2$
and
$\nu_3$ are mixtures of $\nu_\mu$ and $\nu_s$.
In both cases the decay interaction has to be of the form
\begin{equation}
{\cal{L}}_{int} = g_{24} \ \overline{\nu_{4_{L}}^c} \ \nu_{2_{L}} J + h.c.
\end{equation}
where $J$ is a Majoron field that is dominantly iso-singlet (this avoids any
conflict with the invisible width of the $Z$).  Viable
models for both the above cases can be constructed \cite{valle,joshipura}.
However, case (b) needs additional iso-triplet light scalars which cause
potential problems with Big Bang Nucleosynthesis (BBN), and there is some
preliminary evidence from SuperK against $\nu_\mu$--$\nu_s$ mixing
\cite{kajita}. Hence we
only consider case (a), i.e.\ $\nu_2\to\bar\nu_4 + J$ with $\nu_4\approx
\nu_s$, as implicit in Eq.~(\ref{eq:mixing}).
With this interaction, the $\nu_2$ rest-lifetime is given by
\begin{equation}
\tau_2 = \frac
{16 \pi}{g^2} \cdot \
\frac{m_2}{\delta m^2 (1 + x)^2},
\end{equation}
where $\delta m^2 = m^2_2 - m^2_4$ and $x=m_4/m_2 \   (0 < x <1)$.
{}From the value of $\alpha^{-1} = \tau_2/m_2 = 63$~km/GeV found in the fit
and for $x=0$, we
have
\begin{equation}
g^2 \delta m^2 \simeq 0.16\rm\, eV^2
\end{equation}
Combining this with the bound on $g^2$ from $K \rightarrow \mu$ decays of $g^2
< 2.4\times10^{-4}$ \cite{barger1} we have
\begin{equation}
\delta m^2 > 650 \rm\ eV^2 \,.
\end{equation}
Even with a generous interpretation of the uncertainties in the fit,
this $\delta m^2$ implies a minimum mass difference in the range of about
25~eV.
Then $\nu_2$ and $\nu_3$ are nearly degenerate with masses
$\stackrel{\sim}{>} {\cal O}$(25~eV) and $\nu_4$ is relatively light. We assume
that
a similar coupling of $\nu_3$ to $\nu_4$ and J is somewhat weaker
leading to a significantly longer lifetime for $\nu_3$, and the instability of
$\nu_3$ is irrelevant for the analysis of the atmospheric neutrino
data.

For the atmospheric neutrinos in SuperK, two kinds of tests have been proposed
to distinguish between $\nu_\mu$--$\nu_\tau$ oscillations and
$\nu_\mu$--$\nu_s$ oscillations. One is based on the fact that matter effects
are present for $\nu_\mu$--$\nu_s$ oscillations\cite{bdppw} but are nearly absent for
$\nu_\mu$--$\nu_\tau$ oscillations\cite{panta} leading to differences in the zenith angle 
distributions  due to
matter effects on upgoing neutrinos \cite{smirnov}.
The other is the fact that the neutral current rate
will be affected in $\nu_\mu$--$\nu_s$ oscillations but not for
$\nu_\mu$--$\nu_\tau$ oscillations as can be measured in  events
with single $\pi^0$'s \cite{vissani}. In these tests our decay scenario will
behave
as a hybrid in that there is no matter effect but there is some effect in
neutral current rates.

\medskip
\noindent{Long-Baseline Experiments}

The survival probability of $\nu_\mu$ as a function of $L/E$ is given in
Eq.~(1). The conversion probability into $\nu_\tau$ is given by
\begin{equation}
P(\nu_\mu\to\nu_\tau) = \sin^2\theta \cos^2\theta (1-e^{-\alpha L/2E})^2 \,.
\end{equation}
This result differs from $1-P(\nu_\mu\to\nu_\mu)$ and hence is different from
$\nu_\mu$--$\nu_\tau$ oscillations. Furthermore, $P(\nu_\mu\to\nu_\mu)
+ P (\nu_\mu\to \nu_\tau)$ is
not 1 but is given by
\begin{equation}
P (\nu_\mu\to\nu_\mu) + P(\nu_\mu\to\nu_\tau) = 1 - \cos^2\theta (1 -
e^{-\alpha L/E})
\end{equation}
and determines the amount by which the predicted neutral-current rates are
affected compared to the no oscillations (or the $\nu_\mu$--$\nu_\tau$
oscillations) case.
In Fig.~3 we give the results for $P(\nu_\mu\to\nu_\mu)$,
$P(\nu_\mu\to\nu_\tau)$ and $P(\nu_\mu\to\nu_\mu) +
P(\nu_\mu\to\nu_\tau)$ for the decay model and compare them to the
$\nu_\mu$--$\nu_\tau$ oscillations, for both the K2K\cite{who} and
MINOS\cite{minos} (or the corresponding European project\cite{NGS})
long-baseline experiments, with the oscillation and decay parameters as
determined in the fits above.

The K2K experiment, already underway, has a low energy beam $E_\nu
\approx 1\mbox{--}2$~GeV and a baseline $L=250$~km.  The MINOS experiment will have
3 different beams, with average energies $E_\nu = 3,$ 6 and 12 GeV and a
baseline $L=732$~km.  The approximate $L/E_\nu$ ranges are thus 125--250~km/GeV for
K2K and 50--250~km/GeV for MINOS.  The comparisons in Figure 3 show that the
energy dependence of $\nu_\mu$ survival probability and the neutral
current rate can both distinguish between the decay and the oscillation
models.  MINOS and the European project may also have $\tau$ detection
capabilities that would allow additional tests.

\medskip
\noindent{Big Bang Nucleosynthesis}

The decay of $\nu_2$ is sufficiently fast that all the neutrinos ($\nu_e,
\nu_\mu, \nu_\tau, \nu_s$) and the Majoron may be expected to equilibrate in
the early universe before the primordial neutrinos decouple. When they achieve
thermal equilibrium each Majorana neutrino contributes $N_\nu = 1$ and the
Majoron contributes $N_\nu = 4/7$ \cite{kolb}, giving and effective number of
light
neutrinos $N_\nu = 4{4\over7}$ at the time of Big Bang Nucleosynthesis. From
the observed primordial abundances of $^4$He and $^6$Li, upper limits on
$N_\nu$ are inferred, but these depend on which data are
used\cite{olive,lisi,burles}. Conservatively, the upper limit to $N_\nu$ could
extend up to 5.3 (or even to 6 if $^7$Li is depleted in halo stars\cite{olive}).

\medskip
\noindent{Cosmic Neutrino Fluxes}

Since we expect both $\nu_2$ and $\nu_3$ to decay, neutrino beams
from distant sources (such as Supernovae, active galactic nuclei and
gamma-ray bursters) should contain only $\nu_e$ and $\bar\nu_e$
but no $\nu_\mu$, $\bar\nu_\mu$, $\nu_\tau$ and $\bar\nu_\tau$.
This is a very strong prediction of our decay scenario.
We can compare the very different expectations for neutrino flavor mixes
from very distant sources such as AGN's or GRB's. Let us suppose that at the
source the flux ratios are typical of a beam dump, a reasonable assumption:
$N_{\nu_{e}}:N_{\nu_{\mu}}:N_{\nu_{\tau}} = 1:2:0$.  Then, for the
conventional oscillation scenario, when all the $\delta m^2$'s satisfy
$\delta m^2 \ L/4E >>1$, it turns out curiously enough that for
a wide 
variety of choices of neutrino mixing matrices, the final flavor mix is the
same, namely: $N_{\nu_{e}}:
N_{\nu_{\mu}}:N_{\nu_{\tau}} = 1:1:1$.  In the  case of the decay B
scenario, as mentioned here, we have
$N_{\nu_{e}}:N_{\nu_{\mu}}:N_{\nu_{\tau}} = 1:0:0.$ The two are quite distinct.  Techniques for
determining these flavor mixes in future KM3 neutrino telescopes have been
proposed\cite{learned}.

\medskip
\noindent{Reactor and Accelerator Limits}

The $\nu_e$ is essentially decoupled  from the decay state $\nu_2$ so the null
observations from the CHOOZ reactor are satisfied\cite{chooz}. The mixings of $\nu_\mu$ and
$\nu_\tau$ with $\nu_s$ and $\nu_e$ are very small, so there is no conflict
with stringent accelerator limits on flavor oscillations with large $\delta
m^2$~\cite{zuber}.

\medskip
\noindent{Summary}

 In summary, neutrino decay remains a viable
 alternative to neutrino oscillations as an explanation of the atmospheric
 neutrino anomaly. The model consists of two nearly degenerate mass
 eigenstates $\nu_2$, $\nu_3$ with mass separation $\stackrel{\sim}{>} {\cal O}
(25$~eV) from
 another nearly degenerate pair $\nu_1$, $\nu_4$.
The $\nu_\mu$ and $\nu_\tau$ flavors are approximately composed of
$\nu_2$ and $\nu_3$, with a mixing angle $\theta_{23} \simeq 57^\circ$.
The state $\nu_2$ is unstable, decaying to $\bar{\nu}_{4}$ and a Majoron
with a lifetime $\tau_2 \sim 10^{-12}$ sec.  The electron neutrino
$\nu_e$ and a sterile neutrino $\nu_s$ have negligible mixing with $\nu_\mu,
\nu_\tau$ and are approximate mass eigenstates ($\nu_e \approx \nu_1,
\nu_s \approx \nu_4)$, with a small mixing angle $\theta_{14}$ and a
$\delta m_{41}^2 \approx10^{-5}\rm\,eV^2$ to explain the solar
neutrino anomaly.
The states $\nu_3$ and $\nu_4$ are also unstable, but with $\nu_3$ lifetime
somewhat longer and $\nu_4$ lifetime much longer than the
$\nu_2$ lifetime.
This decay
scenario is difficult to distinguish from oscillations because of the
smearing in both L and $E_\nu$ in atmospheric neutrino events.  However,
long-baseline experiments, where $L$ is fixed, should be able to establish
whether the dependence of $L/E_\nu$ is exponential or sinusoidal. In
this scenario only $\nu_1$ is stable.  Thus, neutrinos of supernovae
or of extra galactic origin would be almost entirely $\nu_e$.
The contribution of the electron neutrinos and the Majorons to the cosmological
energy density $\Omega$ is negligible and  not relevant for large
scale structure formation.

\section{Conclusion}

In general, when neutrinos have masses and do mix; they can decay as
well as oscillate.  Some neutrino anomalies may be caused by one or the
other or both.  Non-oscillatory explanations have to be ruled out
experimentally.  Eventually, invisible neutrino decays can be severely
constrained when data from Long Baseline experiments with good
resolution in L and E become available.

\section*{Acknowledgements}

I thank the organizers for the invitation to give this talk and for
selecting the title.  I thank all my collaborators:  A. Acker,
V. Barger, J. G. Learned, P. Lipari, M. Lusignoli and T.J. Weiler.  This
work is partially supported by the U.S.D.O.E. under grant DOE-FG-03-94ER40833.


\begin{figure}[htp]
\centerline{\epsfysize=3.5truein\epsfbox{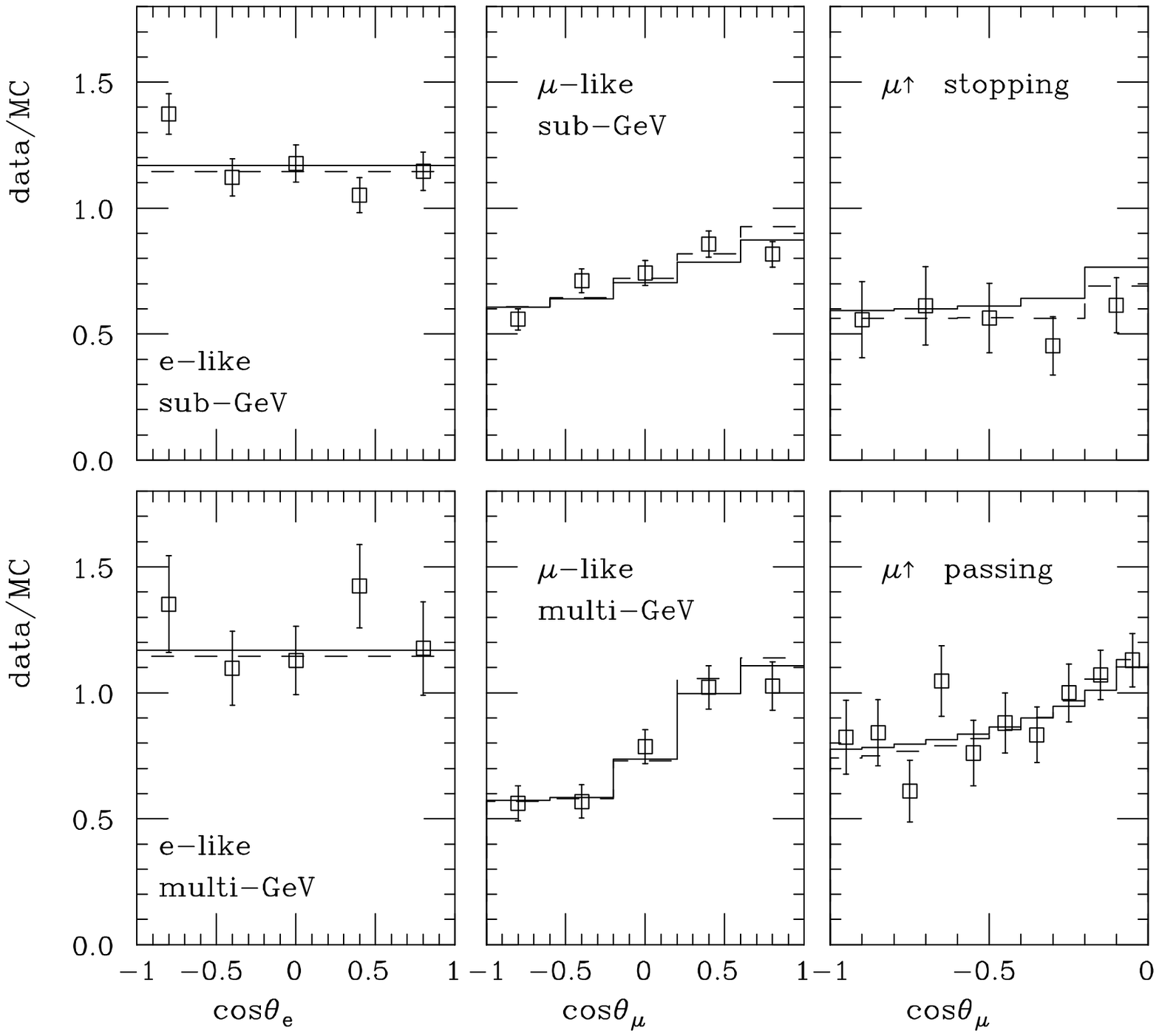}}
\caption{Comparison of decay model (solid histograms) and
$\nu_\mu$--$\nu_\tau$ oscillation model (dashed histograms) with SuperK data.}
\end{figure}

\begin{figure}[htp]
\centerline{\epsfysize 4.5 truein
\epsfbox{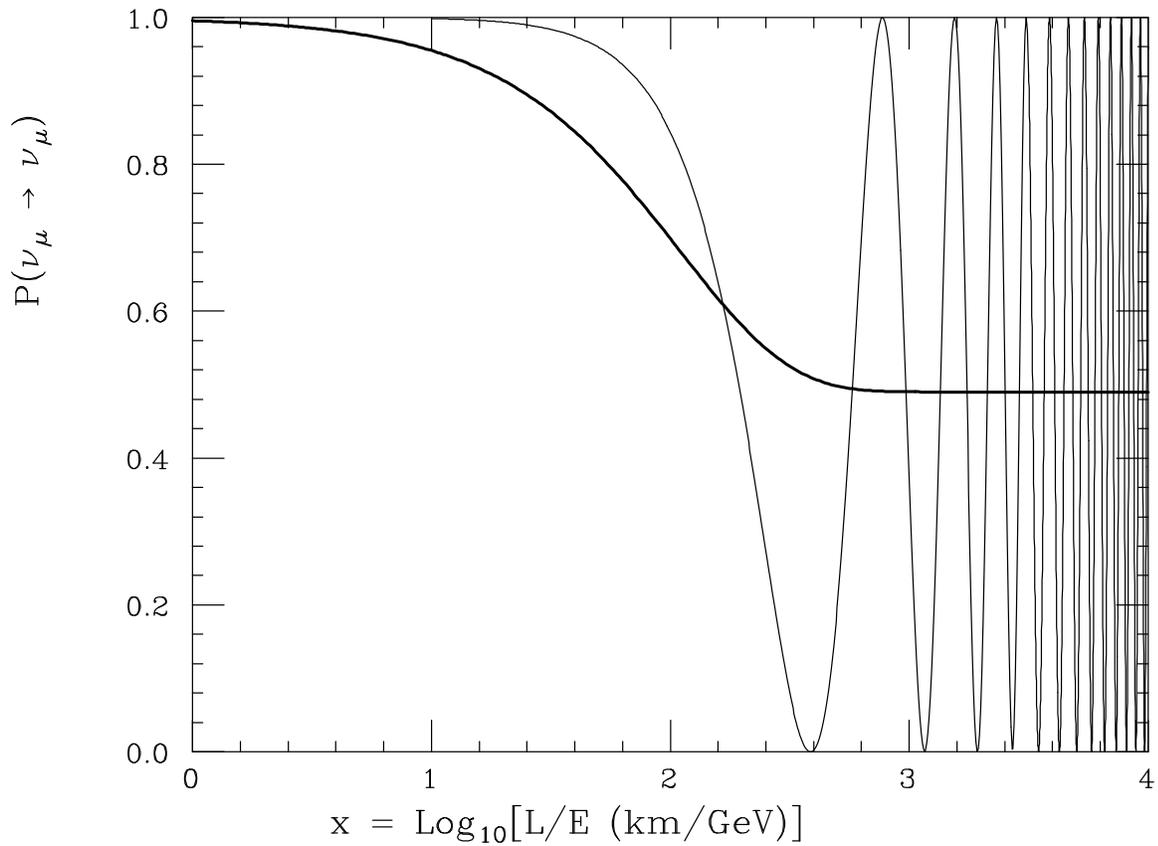}}
%
%
\caption{Survival probabiliity for $\nu_\mu$ versus $\log_{10}(L/E)$ for the
decay model (heavy solid curve) and $\nu_\mu$ oscillation model (thin curve).}
\end{figure}

%
%

\begin{figure}
\centering\leavevmode
\epsfxsize=5.5in\epsffile{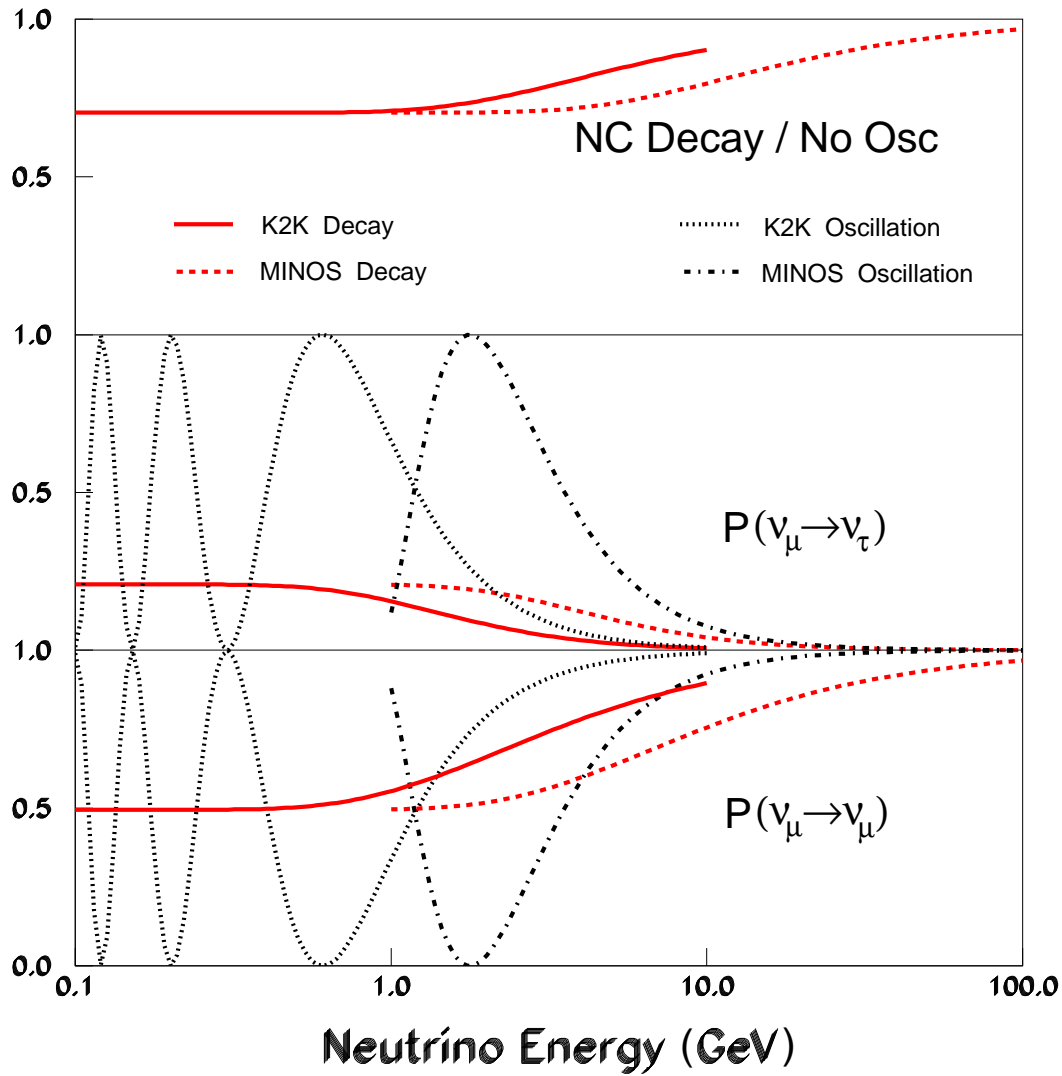}
\vspace{.5in}

\caption[]{Long-baseline expectations for the K2K and MINOS long-baseline
experiments from the
decay model and the $\nu_\mu$--$\nu_\tau$ oscillation model. The upper panel
gives the
neutral current predictions compared to no oscillations (or
$\nu_\mu$--$\nu_\tau$ oscillations).}
\end{figure}

\end{document}